\documentclass{appolb}
\usepackage{epsfig}

\begin{document}
\title{A Mass Formula from Light to Hypernuclei }
{
\author{ C. Samanta$^{1,3}$ \thanks{Presented at the XXIX Mazurian Lakes Conference on Physics, Piaski, Poland,
August 30 - September 6, 2005} \thanks{E-mail:chhanda.samanta@saha.ac.in; csamanta@vcu.edu}, P. Roy Chowdhury$^1$, 
D. N. Basu$^2$ \address{$^1$Saha Institute of Nuclear Physics, 1/AF Bidhan Nagar, Kolkata 700 064, India. \\$^2$Variable  Energy  Cyclotron  Centre,  1/AF Bidhan Nagar, Kolkata 700 064, India.\\$^3$Phys. Dept., Virginia Commonwealth Univ., Richmond, VA 23284-2000, U.S.A.}
}
\maketitle
\begin{abstract}

Simultaneous description of ordinary and hypernuclei masses by a
single mass formula has been a great challenge in nuclear physics.
Hyperon-separation energies of about forty Lambda($\Lambda$), three Lambda-Lambda($\Lambda\Lambda$),
one Sigma($\Sigma$) and seven Cascade($\Xi$) hypernuclei have been experimentally
found. Many of these nuclei are of light masses. We prescribe a new
mass formula, called BWMH, which describes the normal and hypernuclei
on the same footing. It is based on the modified-Bethe-Weizs\"acker mass
formula (BWM). BWM is basically an extension of the Bethe-Weizs\"acker
mass formula (BW) for light nuclei. The parameters of BWM were
optimized by fitting about 3000 normal nuclei available recently. The
original Bethe-Weizs\"acker mass formula (BW) was designed for medium
and heavy mass nuclei and it fails for light nuclei. Two earlier works
on hypernuclei based on this BW show some limitations. The BWMH gives
improved agreement with the experimental data for the line of
stability, one-neutron separation energy versus neutron number spectra
of normal nuclei, and the hyperon-separation energies from
hypernuclei. The drip lines are modified for addition of a $\Lambda$
hyperon in a normal nucleus.


\end{abstract}

\section{Introduction}
\label{section1}

In the last two decades scientific activities have been focussed on hypernuclear physics which is at the boundary between nuclear and particle physics. To obtain a comprehensive view of the basic properties and fundamental interactions of the hadronic system many sophisticated experiments have been done in these interdisciplinary fields. Separation energies have been determined for the ground states of about 40 $\Lambda$ hypernuclei, three double-$\Lambda$ hypernuclei \cite{Ba90}, \cite{Ta01}, several $\Xi^-$(S=-2) hypernuclei \cite{Ba90} and only one bound $\Sigma$ hypernucleus. Several studies suggest that due to strongly repulsive $\Sigma$-nucleus potential $\Sigma$'s are unbound in nuclei, except for the very special case of nuclei with mass number A=4 \cite{Saha04}. No bound state data exists on the  $\Theta^+$ (S=+1, mass $\sim$1530 MeV, width $<$15 MeV ) hypernucleus with exotic pentaquark existence of which was predicted in 1997 \cite{Di97} and announced in 2003 at Spring-8, Japan \cite{Na03} along with several claims of nonexistences \cite{Cl05}. Calculations in a relativistic mean-field formalism (RMF) suggest that as there is an attractive $\Theta^+$-nucleus interaction, the $\Theta^+$ particle can be bound in nuclei and, the $\Theta^+$  hypernuclei would be bound more strongly than $\Lambda$ hypernuclei \cite{Zh05}. Searches for more experimental data on bound $\Theta^+$ hypernuclei are on for a large number of hypernuclei, including   $\Lambda$, $\Lambda\Lambda$, $\Xi$ and $\Sigma$ hyperons. A single mass formula \cite{arxiv} for both the strange and nonstrange nuclei has been formulated to predict the binding energies of all of them on the same footing. Its predictions compare well with the available experimental data. It is not applicable for repulsive potential. 

\section{Generalised mass formula for non-strange and strange nuclei}
\label{section2}

Generalization of mass formula was pursued starting from the
modified-Bethe-Weizs\"acker mass formula (BWM) preserving the normal
nuclear matter properties\cite{Bc04}. The BWM is basically the
Bethe-Weizs\"acker mass formula extended for light nuclei
\cite{Sa02} which can explain
the gross properties of binding energy versus nucleon number curves
of all non- strange normal nuclei from Z=3 to Z=83. A systematic search of experimental data of hyperon separation energy ($S_Y$) for $\Lambda$, $\Lambda \Lambda$, $\Sigma^0$ and $\Xi^-$
hypernuclei leads to a generalised mass formula (BWMH) for hyper and
non-strange nuclei. The hypernucleus is considered as a core of
normal nucleus plus the hyperon(s). Strangeness and hyperon-mass
dependent terms are explicitly included in BWMH breaking the
$SU_F(3)$ symmetry and the binding energy is given as

\begin{eqnarray}
B(A,Z) = &&15.777A-18.34A^{2/3}-0.71\frac{Z(Z-1)}{A^{1/3}}-\frac{23.21(N-Z_c)^2}{[(1+e^{-A/17})A]}\nonumber\\
              &&+(1-e^{-A/30})\delta+ n_Y [0.0335(m_Y) - 26.7 - 48.7 \mid S \mid}{A^{-2/3}], \nonumber\\
\label{seqn1}
\end{eqnarray}
\noindent 
where $\delta=12A^{-1/2}$ for $N,Z_c$ even, $=-12A^{-1/2}$ for $N,Z_c$ odd, = 0 otherwise, $n_Y$ = number of hyperons in a nucleus, $m_Y$ = mass of the hyperon in $MeV$, $S$ = strangeness of the hyperon and
mass number $A = N + Z_c + n_Y$ is equal to the total number of
baryons. $N$ and $Z_c$ are the number of neutrons and protons
respectively while the $Z$ in eqn.(1) is  given by $Z = Z_c + n_Y q$
where $q$ is the charge number (with proper sign) of
hyperon(s) constituting the hypernucleus. For non-strange (S=0)
normal nuclei, $Z_c = Z$ as $n_Y$ =0. The choice of $\delta$
value depends on the number of neutrons and protons in both normal
and hypernuclei. For example, in case of $^{9}_{\Lambda}Li$,
$\delta=-12A^{-1/2}$ as the (N, $Z_c$) combination is odd-odd,
whereas, for non-strange normal $^{9}Li$ nucleus $\delta=0$ for
A=9(odd).

The hyperon separation energy $S_Y$ defined as

\begin{equation}
S_Y = B(A,Z)_{hyper} - B(A-n_Y, Z_c)_{core},
\label{seqn2}
\end{equation}
\noindent
is the difference between the binding energy of a hypernucleus and the binding energy of
its non-strange core nucleus. On the other hand, the separation energies of single neutron ($S_n$) and proton ($S_p$) from the hypernuclei containing single $\Lambda$ inside the nucleus are defined as
\begin{equation}
S_n = B(A,Z)_{hyper} - B(A-1, Z)_{hyper}, ~S_p = B(A,Z)_{hyper} - B(A-1, Z-1)_{hyper}.
\label{seqn3}
\end{equation}
\noindent

It is interesting to note that the
values of $S_Y$ using eqn.(2) are in reasonable agreement with the available
experimental data of all known bound hypernuclei. Fig.[1] shows
plots of $S_Y$ versus A for $\Lambda$ and $\Lambda \Lambda$, $\Xi^-$
 and $\Theta^+$ hypernuclei. BWMH predictions of the first three are in good agreement with the available experimental data
\cite{Ba90,Ta01} and the same for the last one are in close
agreement with the quark mean field (QMF) calculations \cite{Sh05}. Available experimental binding energy values for
$\Sigma^+$ binding energy in $^4_\Sigma He$ are $4.4 \pm 0.3 \pm 1
MeV$ \cite{Naga98}, $2.8 \pm 0.7 MeV$ \cite{Outa94}, $4 \pm 1 MeV$
\cite{Haya92}. The BWMH predicts binding energy of $\Sigma^0$ ($m_Y
= 1192.55 MeV$) and $\Sigma^+$ ($m_Y = 1189.37 MeV$) in
$^4_{\Sigma^0}He$ (=$\Sigma^0$ + $^3_2 He$) as $2.69 MeV$ and
$^4_{\Sigma^+}He$ (=$\Sigma^+$ + $^3_1 H$) as $1.6 MeV$
respectively. Search for bound $\Sigma$ hypernuclei has led to the
conclusion that a $\Sigma$-nucleus potential is strongly repulsive
\cite{Saha04} excepting $^4_\Sigma He$ and without changing any parameter BWMH
reproduces the binding energy of the $^4_\Sigma He$ hypernucleus.
For  $\Sigma^{0} + ^2H$, $\Sigma^{+} + ^2H$, and $\Sigma^{-} + ^2H$
hypernuclei the separation energies predicted by BWMH are -3.53 MeV,
-4.62 MeV and -3.37 MeV respectively indicating that these light
Sigma hypernuclei would be unbound, even if the potential is
attractive. So far no bound state of these hypernuclei could be
found in the experiment. These observations suggests that further
data on $\Sigma$-hypernuclei are necessary to determine more
conclusively whether the $\Sigma$ feels attraction or repulsion.

The effect of addition of a single $\Lambda$ in a non-strage normal nucleus can be seen through the one-neutron and one-proton separation energies tabulated in Table~1. Since hypernuclei are more bound than normal nuclei as a result of increase of nuclear potential depth, drip lines for single $\Lambda$ hypernuclei spread out on the either sides. This observation may have important consequences in the astrophysical objects where bound hypernuclei may exist and form strange stars.

Earlier, Dover and Gal \cite{Do93} prescribed two separate mass
formulae for $\Lambda$ and $\Xi$ hypernuclei by introducing several
volume and symmetry terms in Bethe-Weizs\"acker mass formula (BW), but it did not reproduce experimental data.
Mass formula proposed by Levai et al. \cite{Le98} gave reasonable description of
the experimental data on $\Lambda$ and $\Lambda\Lambda$ hyperon(s)
separation energies, but the binding energy per nucleon diverges as mass number $A$ goes to
infinity. Since none of them contain explicit hyperon
mass in their formulae, they can not be used for binding energy
calculation of other hypernuclei. BWMH is not plagued with such divergences and
nuclear saturation properties are well preserved for large $A$.
\begin{table}
\caption{One-nucleon separation energies on driplines for each element with the lowest and highest number of bound neutrons in normal [17] and $\Lambda$-hypernuclei.}

\begin{tabular}[pos]{cccccccccc}
\hline
\hline
At.&Normal&Normal&Hyper&Hyper&At.&Normal&Normal&Hyper&Hyper \\
No.&p-drip&n-drip&p-drip&n-drip&No.&p-drip&n-drip&p-drip&n-drip \\
\hline
Z&$N,S_p$&$N,S_n$&$N,S_p$&$N,S_n$&Z&$N,S_p$&$N,S_n$&$N,S_p$&$N,S_n$ \\ \hline
\hline
  3&  2,   3.36&  8,    .58&  1,   1.22&  8,   1.87&  4&  2,   1.11& 10,    .90&  2,   3.79& 10,   1.84\\
  5&  3,    .74& 12,    .99&  3,   2.39& 12,   1.73&  6&  3,    .17& 14,   1.01&  3,   1.80& 14,   1.63\\
  7&  5,   1.98& 16,    .97&  4,    .24& 16,   1.50&  8&  5,   1.98& 18,    .94&  4,    .44& 18,   1.40\\
  9&  6,    .09& 20,    .89&  6,    .81& 22,    .01& 10&  6,    .64& 22,    .87&  6,   1.39& 24,    .04\\
 11&  8,    .54& 24,    .84&  8,   1.05& 26,    .08& 12&  8,   1.37& 26,    .84&  7,    .05& 28,    .13\\
 13& 10,    .71& 28,    .84& 10,   1.10& 30,    .18& 14&  9,    .07& 30,    .86&  9,    .56& 32,    .24\\
 15& 12,    .74& 34,    .04& 12,   1.06& 34,    .31& 16& 11,    .46& 36,    .12& 11,    .86& 36,    .37\\
 17& 14,    .68& 38,    .21& 14,    .95& 38,    .44& 18& 13,    .69& 40,    .29& 13,   1.02& 40,    .51\\
 19& 16,    .57& 42,    .37& 16,    .80& 42,    .58& 20& 15,    .81& 44,    .45& 15,   1.08& 46,    .07\\
 21& 18,    .42& 46,    .53& 18,    .62& 48,    .17& 22& 17,    .84& 50,    .07& 17,   1.08& 50,    .25\\
 23& 20,    .24& 52,    .17& 20,    .42& 52,    .34& 24& 19,    .81& 54,    .25& 19,   1.02& 54,    .41\\
 25& 22,    .04& 56,    .34& 22,    .20& 58,    .05& 26& 21,    .73& 58,    .41& 21,    .92& 60,    .13\\
 27& 25,    .76& 62,    .08& 25,    .89& 62,    .22& 28& 23,    .61& 64,    .16& 23,    .78& 64,    .29\\
 29& 27,    .48& 66,    .24& 27,    .60& 68,    .00& 30& 25,    .47& 68,    .31& 25,    .62& 70,    .08\\
 31& 29,    .19& 72,    .04& 29,    .30& 72,    .16& 32& 27,    .30& 74,    .11& 27,    .44& 74,    .22\\
 33& 32,    .66& 76,    .18& 31,    .00& 76,    .30& 34& 29,    .11& 78,    .25& 29,    .23& 80,    .05\\
 35& 34,    .33& 82,    .02& 34,    .42& 82,    .12& 36& 32,    .62& 84,    .08& 31,    .02& 84,    .19\\
 37& 36,    .00& 86,    .15& 36,    .08& 86,    .25& 38& 34,    .36& 88,    .21& 34,    .46& 90,    .04\\
 39& 39,    .33& 92,    .01& 39,    .40& 92,    .10& 40& 36,    .10& 94,    .07& 36,    .19& 94,    .16\\
 41& 42,    .58& 96,    .13& 41,    .04& 96,    .22& 42& 39,    .45& 98,    .19& 39,    .53&100,    .03\\
 43& 44,    .20&102,    .01& 44,    .26&102,    .09& 44& 41,    .14&104,    .07& 41,    .22&104,    .15\\
 45& 47,    .40&106,    .13& 47,    .45&106,    .20& 46& 44,    .39&108,    .18& 44,    .46&110,    .03\\
 47& 49,    .01&112,    .02& 49,    .06&112,    .09& 48& 46,    .06&114,    .07& 46,    .12&114,    .14\\
 49& 52,    .14&116,    .12& 52,    .18&116,    .19& 50& 49,    .25&118,    .17& 49,    .30&120,    .04\\
 51& 55,    .24&122,    .03& 55,    .28&122,    .09& 52& 52,    .38&124,    .07& 52,    .43&124,    .14\\
 53& 58,    .30&126,    .13& 58,    .33&128,    .00& 54& 54,    .02&128,    .17& 54,    .07&130,    .05\\
 55& 61,    .34&132,    .04& 61,    .37&132,    .10& 56& 57,    .11&134,    .08& 57,    .16&134,    .14\\
 57& 64,    .35&136,    .13& 64,    .38&138,    .02& 58& 60,    .17&140,    .01& 60,    .21&140,    .06\\
 59& 67,    .35&142,    .05& 67,    .37&142,    .11& 60& 63,    .21&144,    .10& 63,    .25&144,    .15\\
 61& 70,    .32&146,    .14& 70,    .34&148,    .04& 62& 66,    .22&150,    .03& 66,    .25&150,    .08\\
 63& 73,    .28&152,    .07& 73,    .30&152,    .13& 64& 69,    .22&154,    .11& 69,    .25&156,    .01\\
 65& 76,    .22&158,    .01& 76,    .24&158,    .06& 66& 72,    .19&160,    .05& 72,    .22&160,    .10\\
 67& 79,    .15&162,    .09& 79,    .17&162,    .14& 68& 75,    .15&164,    .13& 75,    .18&166,    .04\\
 69& 82,    .07&168,    .03& 82,    .08&168,    .08& 70& 78,    .10&170,    .07& 78,    .12&170,    .12\\
 71& 86,    .31&172,    .11& 86,    .32&174,    .02& 72& 81,    .03&176,    .02& 81,    .05&176,    .06\\
 73& 89,    .20&178,    .06& 89,    .21&178,    .10& 74& 85,    .27&180,    .10& 85,    .29&182,    .01\\
 75& 92,    .08&184,    .01& 92,    .08&184,    .05& 76& 88,    .17&186,    .04& 88,    .18&186,    .09\\
 77& 96,    .25&188,    .08& 96,    .25&190,    .00& 78& 91,    .07&190,    .12& 91,    .08&192,    .04\\
 79& 99,    .11&194,    .04& 99,    .12&194,    .08& 80& 95,    .24&196,    .07& 95,    .25&196,    .11\\
 81&103,    .25&198,    .11&103,    .25&200,    .03& 82& 98,    .11&202,    .03& 98,    .12&202,    .07\\
 83&106,    .08&204,    .07&106,    .09&204,    .10&   &&&& \\ \hline
\hline
\end{tabular}
\end{table}

\begin{figure}[htbp]
\eject\centerline{\epsfig{file=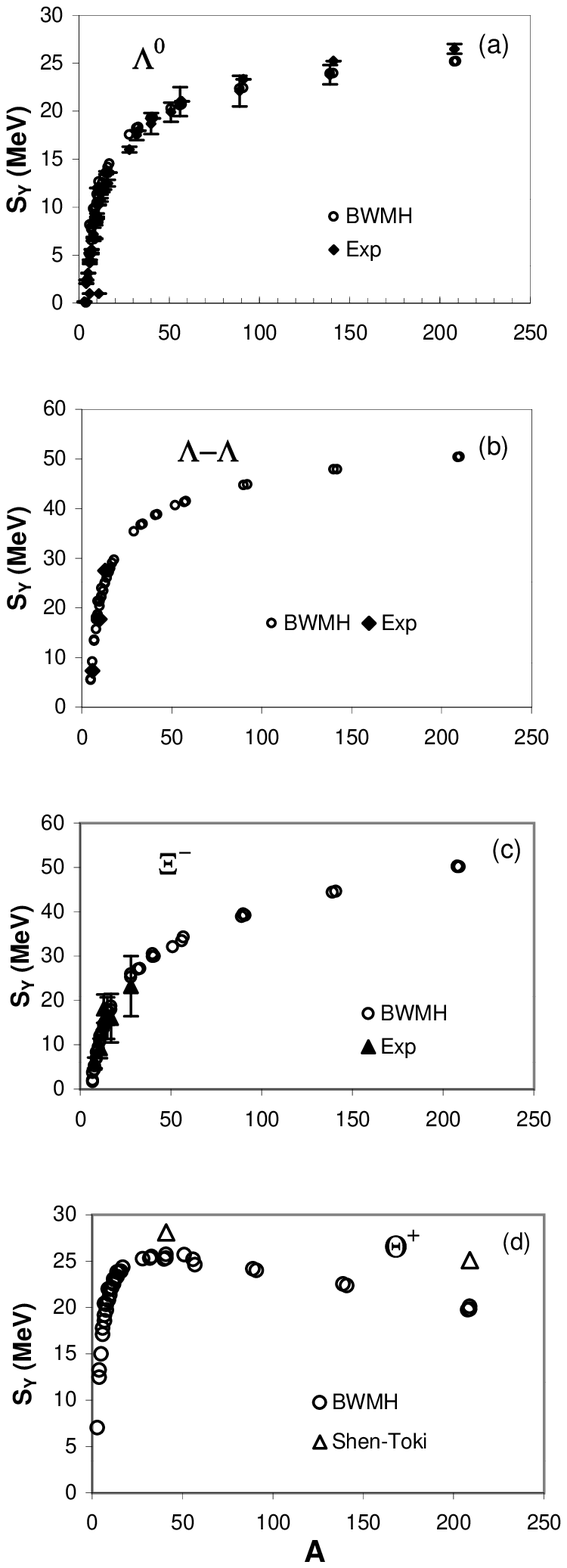,height=20cm,width=8.9cm}}
\caption
{Plots of hyperon-separation energies of (a) $\Lambda$,(b)
$\Lambda\Lambda$, (c) $\Xi^-$ and (d) $\Theta^+$ hypernuclei with
mass number(A). BWMH predictions are compared with experimental data
for $\Lambda$, $\Lambda\Lambda$, $\Xi^-$ and QMF predictions
\cite{Sh05} for the three $\Theta^+$ hypernuclei.}
\label{fig1}
\end{figure}

\section{Summary and Conclusion}
\label{section3}

In summary, a simple one line mass formula (BWMH) applicable to
normal as well as strange hypernuclei is developed by introducing
hyperon mass and strangeness dependent SU(6) symmetry breaking terms
in BWM. Since it is
not applicable for repulsive potential it does not predict negative
sigma separation energy for heavier $\Sigma$-nuclei. It
predicts that $\Theta$ hypernuclei would be more strongly bound than
$\Lambda$ hypernuclei which is in good agreement with the quark mean
field calculation \cite{Sh05}. Calculations of $S_p$ and $S_n$ of
normal nuclei and nuclei with single $\Lambda$ hyperon inside the
nucleus indicate that due to stronger $\Lambda$-nucleon interaction, the mean field potential gets
modified by addition of single $\Lambda$ hyperon to the core.
Introduction of $\Lambda$ inside a nucleus thus alters the usual
neutron and proton driplines and gives birth of new nuclei of
astrophysical interest beyond normal neutron and
proton driplines.

\end{document}